\documentclass[12pt]{article}
\usepackage{rotating}
\usepackage{longtable}
\usepackage{amsmath}
\usepackage{lscape}
\usepackage{setspace}
\usepackage{graphicx}

\usepackage{verbatim, color, amssymb, epsfig}

\def\bSig\mathbf{\Sigma}

\def\log{\hbox{log}}

\def\dfrac#1#2{{\displaystyle{#1\over#2}}}

\def\boxit#1{\vbox{\hrule\hbox{\vrule\kern6pt
 \vbox{\kern6pt#1\kern6pt}\kern6pt\vrule}\hrule}}

\def\cov{\hbox{cov}}

\def\bse{\begin{eqnarray*}}
\def\ese{\end{eqnarray*}}
\def\be{\begin{eqnarray}}
\def\ee{\end{eqnarray}}
\def\bq{\begin{equation}}
\def\eq{\end{equation}}
\def\bse{\begin{eqnarray*}}
\def\ese{\end{eqnarray*}}

\def\wh{\widehat}

\def\bea{\begin{eqnarray}}
\def\eea{\end{eqnarray}}
\def\bd{\begin{displaymath}}
\def\ed{\end{displaymath}}
\def\bda{\begin{eqnarray*}}
\def\eda{\end{eqnarray*}}

\def\diag{\mbox{Diag}}

\def\pb{\mbox{P}^{\beta}}
\def\bv{b^{\nu}}
\def\bb{b^{\beta}}
\def\wsw{(W^T{\Sigma^*}^{-1}W+\pb)^{-1}}

\def\nn{\nonumber}

\newcommand{\uiota}             {\mbox{\boldmath$\uiota$}}

\newtheorem{lemma}{\noindent\mbox{Lemma}}
\newtheorem{theorem}{\noindent\mbox{Theorem}}
\newtheorem{corollary}{\noindent\mbox{Corollary}}

\def\btheorem{\begin{theorem}\emph{}\def\etheorem{\end{theorem}}}

\begin{document}

\begin{center}
\textbf{Uncertainty Estimation in Functional Linear Models}\\
\end{center}

\vspace{0.25cm}

\begin{center}
\today
\end{center}

\vspace{0.25cm}

\begin{center}
Tapabrata Maiti\footnote{maiti@stt.msu.edu}, Abolfazl Safikhani\footnote{as5012@columbia.edu}, Ping-Shou Zhong\footnote{pszhong@stt.msu.edu}\\
\vspace{0.25cm}
$^{1,3}$Department of Statistics and Probability, Michigan State University, East Lansing, Michigan, U.S.A.\\
$^{2}$Department of Statistics, Columbia University, New York, New York, U.S.A \\
\end{center}

\author{Tapabrata Maiti$^{1,*}$\email{maiti@stt.msu.edu}, Abolfazl Safikhani$^{2,**}$\email{as5012@columbia.edu}
and
Ping-Shou Zhong$^{3,***}$\email{pszhong@stt.msu.edu} \\
$^{1,3}$Department of Statistics and Probability, Michigan State University, East Lansing, Michigan, U.S.A.\\
$^{2}$Department of Statistics, Columbia University, New York, New York, U.S.A \\
%$^{3}$Department of Statistics and Probability, Michigan State University, East Lansing, Michigan, U.S.A.
}

%
%\center{\textbf{Prediction Error Estimation of Mixed Effects in Functional Linear Models}}

%\bf\Large{ Prediction Error Estimation of Mixed Effects in Functional Linear Models}
\date{}
%\maketitle

\textbf{Abstract} Functional data analysis is proved to be useful in many scientific applications. The physical process is observed as curves and often there are several curves observed due to multiple subjects, providing the replicates in statistical sense. The recent literature develops several techniques for registering the curves and associated model estimation. However, very little has been investigated for statistical inference, specifically uncertainty estimation. In this article, we consider functional linear mixed modeling approach to combine several curves. We concentrate measuring uncertainty when the functional linear mixed models are used for prediction. Although measuring the uncertainty is paramount interest in any statistical prediction, there is no closed form expression available for functional mixed effects models. In many real life applications only a finite number of curves can be observed. In such situations it is important to asses the error rate for any valid statistical statement. We derive theoretically valid approximation of uncertainty measurements that are suitable along with modified estimation techniques. We illustrate our methods by numerical examples and compared with other existing literature as appropriate. Our method is computationally simple and often outperforms the other methods.
%  This will produce the submission and review information that appears
%  right after the reference section.  Of course, it will be unknown when
%  you submit your paper, so you can either leave this out or put in
%  sample dates (these will have no effect on the fate of your paper in the
%  review process!)

\date{
%{\it Received October} 2007. {\it Revised February} 2008.  {\it
%Accepted March} 2008.
}

%  These options will count the number of pages and provide volume
%  and date information in the upper left hand corner of the top of the
%  first page as in published papers.  The \pagerange command will only
%  work if you place the command \label{firstpage} near the beginning
%  of the document and \label{lastpage} at the end of the document, as we
%  have done in this template.

%  Again, putting a volume number and date is for your own amusement and
%  has no bearing on what actually happens to your paper!

%\volume{}
%\pubyear{}
%\artmonth{}
%
%%  The \doi command is where the DOI for your paper would be placed should it
%%  be published.  Again, if you make one up and stick it here, it means
%%  nothing!
%
%\doi{}

%  This label and the label ``lastpage'' are used by the \pagerange
%  command above to give the page range for the article.  You may have
%  to process the document twice to get this to match up with what you
%  expect.  When using the referee option, this will not count the pages
%  with tables and figures.

%\label{firstpage}

%  put the summary for your paper here

%\begin{abstract}
%
%\end{abstract}

%  Please place your key words in alphabetical order, separated
%  by semicolons, with the first letter of the first word capitalized,
%  and a period at the end of the list.
%

\textbf{keywords}
%\begin{keywords}
Basis functions; B-splines; Bias correction; Estimating equations; Functional mixed models; Karhunen-Lo$\grave{e}$ve expansion;  Prediction interval; Random effects;
%\end{keywords}

%  As usual, the \maketitle command creates the title and author/affiliations
%  display

%\maketitle

%  If you are using the referee option, a new page, numbered page 1, will
%  start after the summary and keywords.  The page numbers thus count the
%  number of pages of your manuscript in the preferred submission style.
%  Remember, ``Normally, regular papers exceeding 25 pages and Reader Reaction
%  papers exceeding 12 pages in (the preferred style) will be returned to
%  the authors without review. The page limit includes acknowledgements,
%  references, and appendices, but not tables and figures. The page count does
%  not include the title page and abstract. A maximum of six (6) tables or
%  figures combined is often required.''

%  You may now place the substance of your manuscript here.  Please use
%  the \section, \subsection, etc commands as described in the user guide.
%  Please use \label and \ref commands to cross-reference sections, equations,
%  tables, figures, etc.
%
%  Please DO NOT attempt to reformat the style of equation numbering!
%  For that matter, please do not attempt to redefine anything!

\section{Introduction}\label{introduction}
Functional data analysis received considerable attention in last couple of decades due to its applicability in various scientific studies (Ramsay and Silverman, 1997, 2010). A functional data consists of several functions or curves that are observed on a sufficiently large number of grid points. The dual characteristics of functional data, namely, maintaining the smoothness of individual curves through repeated observations and combining several curves poses the main challenge in functional data analysis (Morris and Carroll, 2006). In relatively simple situations where the curves could be represented by simple parametric models, this data characteristics can easily be handled by well established parametric mixed models (Laird and Ware, 1982, Verbeke and Molenbergs, 2000, Demidenko, 2013, McCulloch and Searle, 2001, Jiang 2007). In complex biological or physical processes, the parametric mixed models are not appropriate and thus modern functional mixed model techniques are developed. For a complete description and comparisons, instead of repeating the literature, we suggest Guo (2002a), Morris and Carroll (2006), Antoniadis and Sapatinas (2007), Chen and Wang (2011), Liu and Guo (2011), Maiti, Sinha and Zhong (2016) for recent developments in functional mixed linear models and their applications. Some of the related works, but not a complete list includes Brumback and Rice (1998), Goldsmith, Crainiceanu, Caffo and Reich (2011), Greven, Crainiceanu, Caffo and Reich (2011), Krafty, Hall and Guo (2011) and Staicu, Crainiceanu and Carroll (2010).

As noted by Antoniadis and Sapatinas (2007), much of the work done in this area is on estimation and only limited attention has been paid for inference. Morris and Carroll (2006) provided a comprehensive account of functional mixed models from Bayesian perspective. In Bayesian approach, the inference is automatic along with the estimation. However, this is not the case in frequentist perspective. Antoniadis and Sapatinas (2007) developed testing of random effects and fixed effects when a wavelet decomposition is used for smoothing the functional curves. They also  observed the theoretical limitations of the procedures developed for functional mixed models, for example, Guo (2002a,b). Thus research developing theoretically valid inferential techniques for functional mixed models is important.

Prediction is one of the main objectives in functional data analysis in many applications. For example, one may like to predict the number of CD4 cells in HIV patients or thai circumference (related to body weight) for obese children in a future time point. The mixed effect, combination of fixed and random effects are typically used for such prediction. For statistical inference, standard error estimation of the mixed effects is an integral part of data analysis. However, this is a non trivial problem even in simple parametric random effect models. See, Kackar and Harville (1984), Rao and Molina (2015) and Jiang (2007) for various developments. We developed the approximation theory for standard error estimation as a measure of uncertainty  for mixed effects in a functional linear mixed models. Our approach is frequentist. To our knowledge, the prediction error estimation in this sense for a functional mixed model is new. Guo (2002a) reported subject (curve) specific 95\% confidence intervals in his applications. However, like Antoniadis and Sapatinas (2007) we were unable to verify the development with theoretical validation. We believe, Guo's primary goal was in developing the functional mixed models and user friendly estimation rather than prediction interval estimation. This work, in this sense, is complementary to Guo (2002a) and an important contribution to the functional mixed model methodology.

In this paper we consider a functional mixed model similar to Chen and Wang (2011). We also follow the penalized spline smoothing technique similar to them. Of course there are other possible smoothing techniques, such as wavelet (Morris and Carroll, 2006, Antoniadis and Sapatinas (2007),  functional principal components (Yao, M\"{u}ller and Wang, 2005), Kauermann and Wegener, 2011 and Staicu, Crainiceanu and Carroll, 2010). For comparison, contrasts,  computation and further literature review of various smoothing techniques in this context we refer to Chen and Wang (2011). The section \ref{Preliminary} introduces the models and model estimation. The section also develops the approximation theory of prediction error. The simulation study and real data examples are given in Section \ref{numerical study}. We conclude the development in Section \ref{Discussion}. The proofs are deferred to Appendix.

\section{Models and Estimation}\label{Preliminary}
We consider a functional mixed-effect  model as
\begin{equation}
Y(t)=X(t)^T\beta(t)+Z(t)^T\nu(t)+e(t)
\label{FMEM0}
\end{equation}
where $t\in(0,1)$, $\beta(t)$ is a $p$-dimensional fixed coefficient function, $\nu(t)$ is a $q$-dimensional subject-specific random coefficient functions and $e(t)$ is a Gaussian noise process with $e(t)\sim N(0,\sigma_e^2)$ and $\cov(e(t),e(s))=0$. We also assume that $e(t)$ and $\nu(t)$ are independent.

Let the data $\{Y(t_{ij}), X(t_{ij}), Z(t_{ij})\}$ are collected/designed at time points $t_{ij}$ for $i$-th subject, $i=1,\cdots,n$ and $j$-th time point,  $j=1,\cdots,m_i$. Then the  model (\ref{FMEM0})  for this data is
\begin{equation}
Y(t_{ij})=X(t_{ij})^T\beta(t_{ij})+Z(t_{ij})^T\nu_i(t_{ij})+e_{ij}
\label{FMEM}
\end{equation}
where $X(t_{ij})=(X_1(t_{ij}),\cdots, X_p(t_{ij}))^T$ is the covariates, $Z(t)=(Z_1(t),\cdots, Z_q(t))^T$ known design for random effects  $\nu_i(t)=(\nu_{i1}(t),\cdots, \nu_{iq}(t))^T$. These are zero mean Gaussian process with $Cov(\nu_{i}(t), \nu_{i}(s))=\gamma(t,s)$, a $q\times q$ positive definite matrix. We assume $\nu_{ik}(t)$ and $\nu_{jl}(t)$ are independent for $i\neq j$ or $k\neq l$. The objective is to obtain the mean square error for  predicting the mixed effect $A=l^T_0\beta(t)+d^T_0\nu(t)$, where $l_0$ and $d_{0}$ are $p$ and $q$ dimensional known vectors.

Assume that $\beta_k(t), k=1,\cdots, p$ belongs to the normed space of continuous functions with finite second derivatives and the covariance $\gamma(s,t)$ can be decomposed by
$$
\gamma(t,s)=B_{\nu k}(s)^T\Omega_k B_{\nu k} (s).
$$
Then we can approximate $\beta_k(t)$ by
$$
\beta_k(t)=B_k(t)^T\theta_k
$$
where $B_k(t)=(B_{k1}(t), B_{k2}(t),\cdots,B_{kL}(t))^T$ is the vector of $L$ basis functions and $\theta_k=(\theta_{k1},\cdots,\theta_{kL})^T$ is the corresponding coefficients. By Karhunen-Lo$\grave{e}$ve expansion (Ash and Gardner, 1975), the random functions $\nu_{ik}(t)$ can be approximated by
$$
\nu_{ik}(t)=B_{\nu k}(t)^T\alpha_{ik}
$$
where $B_{\nu k}(t)=(B_{\nu k 1}(t),\cdots,B_{\nu k L}(t))^T$, $\alpha_{ik}=(\alpha_{ik1},\cdots,\alpha_{ikL})^T$ and
%$\alpha_{ik}$ are uncorelated random coefficients.
$\alpha_i$ are random coefficients with $\mbox{Var}(\alpha_{ik})=\Omega_k$.  In this paper, we consider B-spline basis functions
for $B_k(t)$ and $B_{\nu k}(t)$. Let $0=\tau_0< \tau_1< \cdots < \tau_{L_0}< \tau_{L_0+1}=1$ be a set of knot points and $\tau_i=\tau_{\min\{\max(i,0),L_0+1\}}$ for any $i=1-r,\cdots, L$, where $L=L_0+r$.  Using these knots, we can define  $L$ normalized B-spline basis function of order $r$. The B-spline basis is defined by
$$
B_{ki}(t)=(\tau_i-\tau_{i-r})[\tau_{i-r},\cdots,\tau_i](\tau-t)_{+}^{r-1}\quad\mbox{for $i=1,\cdots, L$}
$$
where
$$[\tau_{i-r},\cdots,\tau_i]\phi(\tau)=\frac{[\tau_{i-r+1},\cdots,\tau_i]\phi(\tau)-[\tau_{i-r},\cdots,\tau_{i-1}]\phi(\tau)}{\tau_i-\tau_{i-r}}$$
denotes the $r$-th order divided difference for $r+1$ distinct points $\tau_{i-r},\cdots,\tau_i$ of function $\phi$ and
$$ [\tau_{j},\tau_{j+1}]\phi(\tau)=\frac{\phi(\tau_{j+1})-\phi(\tau_{j})}{\tau_{j+1}-\tau_{j}}.$$
 If $\tau_{i}=\tau_{i+1}=\cdots=\tau_{i+m}$, then $[\tau_{i},\cdots,\tau_{i+m}]\phi(\tau)=\frac{\phi^{(m)}(\tau)}{m!}$ for some integer $m$.
We also used B-spline for the $B_{\nu k}(t)$.
%The cubic B-spline is defined as
%\be
%B_{kl,3}(t)=\frac{t-\tau_l}{\tau_{l+3-1}-\tau_{l-1}}B_{kl,2}(t)+\frac{t-\tau_l}{\tau_{l+3-1}-\tau_{l-1}}B_{kl,2}(t)
%\ee

Then the model (\ref{FMEM}) is represented as a linear mixed effect model
\begin{equation}
Y_i=W_i\theta+U_i\alpha_i+e_i
\label{2.3}
\end{equation}
where
$Y_i=(Y(t_{i1}),\cdots, Y(t_{im}))^T_{m\times 1}$, $W_i=(W_{i1}, \cdots, W_{im})^T_{m\times pL}$, $\theta=(\theta_1^T,\theta_2^T,\cdots, \theta_p^T)^T_{pL\times 1}$, $U_i=(U_{i1}, \cdots, U_{im})^T_{m\times qL}$ and $\alpha_i=(\alpha_{i1}^T,\alpha_{i2}^T,\cdots, \alpha_{iq}^T)^T_{qL\times 1}$ with $W_{ij}=(W_{1}(t_{ij})^T, \cdots, W_{p}(t_{ij})^T)^T_{pL\times 1}$
and $U_{ij}=(U_{1}(t_{ij})^T, \cdots, U_{q}(t_{ij})^T)^T_{qL\times 1}$ where $W_{k}(t)=(X_k(t)B_{k1}(t),\cdots, X_k(t)B_{kL}(t))^T_{L\times 1}$ and $U_k(t)=(Z_k(t)B_{\nu k1}(t),\cdots, Z_k(t)B_{\nu kL}(t))^T_{L\times 1}.$  For convenience, denote $R_i=\diag(\sigma_e^2,\cdots,\sigma_e^2)$ as the variance of $e_i=(e_{i1},\cdots,e_{im})^T$. Define $\Omega=\mbox{Diag}\{\Omega_1,\cdots, \Omega_q\}$. It follows that $\mbox{Var}(\alpha_i)=\Omega$. Then,  $\Sigma_i=U_i\Omega U_i^T+R_i$ is the variance of $Y_i$.  Hence, we can write the mixed effect $A$ into a function of $\theta$ and $\alpha$, which is
\be
A=l^T\theta+\sum_{i=1}^nd^T_i\alpha_i=l^T\theta+d^T\alpha
\label{predict-objs}
\ee
where $d=(d_1^T,\cdots,d_n^T)^T$ and
$$l=l_0^T
\left(
\begin{array}{ccc}
B_1^T(t)  & 0  & 0  \\
0  & \ddots  & 0  \\
0  & 0  &  B_p^T(t)
\end{array}
\right)\quad\mbox{and}\quad d_i=d_{i0}^T\left(
\begin{array}{ccc}
B_{\nu 1}^T(t)  & 0  & 0  \\
0  & \ddots  & 0  \\
0  & 0  &  B_{\nu q}^T(t)
\end{array}
\right).
$$

\subsection{Estimation of the fixed parameter and random effects}

Following Chen and Wang (2007), treating the random effect $\alpha_i$ as missing value, define the penalized joint log-likelihood for $Y_i$ and $\alpha_i$ as
\begin{eqnarray}
\ell(\theta,\alpha)&=\sum_{i=1}^n \{(Y_i-W_i\theta-U_i\alpha_i)^TR_i^{-1}(Y_i-W_i\theta-U_i\alpha_i)+\alpha_i^T\Omega^{-1}\alpha_i\}\nn\\
&\quad+\sum_{k=1}^p\lambda_k \theta_k^T\Delta_{\beta k}\theta_k+\sum_{k=1}^q\eta_k\sum_{i=1}^n\alpha_{ik}^T\Delta_{\nu ik}\alpha_{ik}
\label{joint-likelihood}
\end{eqnarray}
where $\Delta_{\beta k}=\int_{0}^{1} B_{k}^{(2)}(t)B^{(2)^T}_{k}(t)dt$, $B_{k}^{(2)}(t)$ is the second derivative of $B_{k}(t)$ with respect to $t$ and $\Delta_{\nu ik}$ are defined similarly based on the B-spline basis functions $B_{\nu k}(t)$.

Then (\ref{joint-likelihood}) can be written as
\begin{eqnarray}
\ell(\theta,\alpha)&=\sum_{i=1}^n \{(Y_i-W_i\theta-U_i\alpha_i)^TR_i^{-1}(Y_i-W_i\theta-U_i\alpha_i)\nn\\
&+\alpha_i^T\Omega^{-1}\alpha_i\} +\theta^T\Delta_{\beta}\theta+\sum_{i=1}^n\alpha_i^T\Delta_{\nu i}\alpha_{i}.
\end{eqnarray}
where $\Delta_{\beta}=\mbox{Diag}\{\lambda_1\Delta_{\beta 1},\cdots, \lambda_p\Delta_{\beta p}\}$ and $\Delta_{\nu i}=\mbox{Diag}\{\eta_1\Delta_{\nu i1},\cdots, \eta_q\Delta_{\nu iq}\}$,

Given $\Omega$, the minimization of $\ell(\theta,\alpha)$ provides
\begin{eqnarray*}
\tilde{\theta}&=(\sum_{i=1}^n W_i^TR_i^{-1}D_{i}W_i+\Delta_{\beta})^{-1} \sum_{i=1}^n W_i^T R_{i}^{-1} D_i Y_i\\
\tilde{\alpha}_i&=(U_i^TR_i^{-1}U_i+\Omega^{-1}+\Delta_{\nu i})^{-1} U_i^T R_{i}^{-1} (Y_i-W_i\tilde{\theta}).
\end{eqnarray*}

%
%which can be written into matrix form
%\begin{eqnarray}
%\tilde{\theta}&=(W^TR^{-1}DW+\Delta_{\beta})^{-1} W^TDY\\
%\tilde{\alpha}_i&=V^{-1}U^TR^{-1}(Y-W\tilde{\theta})
%\end{eqnarray}
%where $W=(W_1^T,W_2^T,\cdots, W_n^T)^T$, $D=\mbox{Diag}(D_1,\cdots, D_n)$, $U=\mbox{Diag}\{U_1,\cdots, U_n\}$ and $V=\mbox{Diag}\{V_1,\cdots, V_n\}$ with $D_i=I-U_i(U_i^TR_i^{-1}U_i+\Omega^{-1}+\Delta_{\nu i})^{-1}U_i^TR_i^{-1}$ and $V_i=U_i^TR_iU_i+\Omega^{-1}+\Delta_{\nu i}$.

By some algebra, we then have
\bea
\label{theta-esti}
\tilde{\theta}&=&(\sum_{i=1}^n W_i^T\Sigma_i^* W_i+\Delta_\beta)^{-1}\sum_{i=1}^n W_i^T{\Sigma_i^*}^{-1}Y_i\\
\tilde{\alpha}&=&\Omega^*U^T{\Sigma^*}^{-1}(Y-W\tilde{\theta})
\label{alpha-esti}
\eea
where $\Sigma_i^*=U_i(\Omega^{-1}+\Delta_{\nu i})^{-1}U_i^T+R_i$ and $\Omega^*=\diag((\Omega^{-1}+\Delta_{\nu 1})^{-1},\cdots, (\Omega^{-1}+\Delta_{\nu n})^{-1})$.

Because $\Sigma^*$ involves some unknown parameters,  we estimate the variance component $\sigma$ in $\Sigma^*$ by solving the following estimating equation
\be
\label{est:equa:variance}
\mathcal{Q}_{VC,k}(\sigma):=-tr\{P\Sigma P\frac{\partial \Sigma^*}{\partial \sigma_k}\}+Y^TP\frac{\partial \Sigma^*}{\partial \sigma_k}PY=0\quad\mbox{for $k=1,\cdots,g$.}
\ee
where $P={\Sigma^*}^{-1}-{\Sigma^*}^{-1}W(W^T{\Sigma^*}^{-1}W+\Delta_\beta)^{-1}W^T{\Sigma^*}^{-1}$. Notice that $\mathcal{Q}_{VC,k}(\sigma)$ contains $g$ estimating equations. Denote the estimate of $\sigma$ by $\hat{\sigma}$.

Note that the above estimating equation is bias-corrected score function {\it from the score function of the restricted log-likelihood function. To see this point, }  we observe that restricted log-likelihood function is the following
\bea
\ell_V(\sigma)&:=&-\log|W{\Sigma^*}^{-1}W|-\log(|\Sigma^*|)-(Y-W\tilde{\theta}-U\tilde{\alpha})^TR^{-1}(Y-W\tilde{\theta}-U\tilde{\alpha})\nn\\
&&-\tilde{\alpha}^T\Omega^{-1}\tilde{\alpha}-\tilde{\theta}^T\Delta_{\beta}\tilde{\theta}-\tilde{\alpha}^T\Delta_{\nu}\tilde{\alpha}\nn\\
&=&-\log|W{\Sigma^*}^{-1}W|-\log(|\Sigma^*|)-(Y-W\tilde{\theta})^T{\Sigma^*}^{-1}(Y-W\tilde{\theta})-\tilde{\theta}^T\Delta_{\beta}\tilde{\theta}\nn\\
&=&-\log|W{\Sigma^*}^{-1}W|-\log(|\Sigma^*|)-Y^TPY.
\eea
Then the derivative of $\ell_V(\sigma)$ is
\be
\frac{\partial\ell_V(\sigma)}{\partial\sigma}=-tr(P\frac{\partial \Sigma^*}{\partial \sigma_k})+Y^TP\frac{\partial \Sigma^*}{\partial \sigma_k}PY
\ee
But $E\{\frac{\partial\ell_V(\sigma)}{\partial\sigma}\}\neq 0$ unless $\Sigma^*=\Sigma$. To make the score equation to be unbiased, we modified the score function $\frac{\partial\ell_V(\sigma)}{\partial\sigma)}$ to be (\ref{est:equa:variance}) such that it is unbiased.

\subsection{Prediction and Prediction Mean Square Error}

A naive prediction of $A$ given in (\ref{predict-objs}) is $\tilde{A}(\sigma)=l'\tilde{\theta}+d'\tilde{\alpha},$ where $\sigma$ is an unknown vector of variance components in $\Sigma$. The $k$-th component of $\sigma$ will be denoted as $\sigma_k$ and $\sigma=(\sigma_1,\cdots,\sigma_g)^T$. Unlike the simple linear mixed models, this prediction is biased as stated in the following theorem.

\btheorem
\label{thm1}
 The prediction is biased and the bias of the prediction of $\tilde{A}(\sigma)$ is
\be
\label{Bias}
%\hspace{-2.5cm}
\mbox{Bias}\{\tilde{A}(\sigma)\}=-l'(W^T{\Sigma^*}^{-1}W+\Delta_{\beta})^{-1}\Delta_{\beta}\theta+d'\Omega^*U^T{\Sigma^*}^{-1}W(W^T{\Sigma^*}^{-1}W+\Delta_{\beta})^{-1}\Delta_{\beta}\theta.
\ee
\etheorem

To reduce the order of bias, we propose a bias-corrected prediction for $A(\sigma)$ as

\begin{eqnarray}
\tilde{A}_c(\sigma)&=l'\tilde{\theta}+d'\tilde{\alpha}+l'(W^T{\Sigma^*}^{-1}W+\Delta_{\beta})^{-1}\Delta_{\beta}\tilde{\theta}\nn \\
&\quad-d'\Omega^*U^T{\Sigma^*}^{-1}W(W^T{\Sigma^*}^{-1}W+\Delta_{\beta})^{-1}\Delta_{\beta}\tilde{\theta}.
\label{bias-corr-prediction}
\end{eqnarray}

The proof is deferred to the Appendix.

\btheorem
\label{thm2}
If $d=(d_1',\cdots,d_n')'$ is sparse, namely only a finite number of $d_j$'s are non-zeros ($j=1,\cdots,n$). Then
the bias of $\tilde{A}_c(\sigma)$ is  of order $n^{-2}$, which is negligible in comparing to $n^{-1}$.
\etheorem

The proof is deferred to the Appendix.
%\vspace*{2ex}

Now we will derive the prediction error formula.
Let $Q=\diag(Q_1,\cdots, Q_n)$ and $D=(W^T\Sigma^{-1}W)^{-1}(I+B(W^T\Sigma^{-1}W)^{-1})^{-1}B(W^T\Sigma^{-1}W)^{-1}$ where $Q_i=\Sigma_i^{-1}U_i\Omega(I+\Delta_{\nu i}\Omega-\Delta_{\nu i}\Omega U_i\Sigma_i^{-1}U_i\Omega)\Delta_{\nu i}\Omega U_i \Sigma_i^{-1}$ and $B=W^TQW+\Delta_\beta$. Define $\Delta_1=(W^T\Sigma^{-1}W)^{-1}W^TQ-DW^T(\Sigma^{-1}+Q)$. Then it can be shown that
\begin{eqnarray}
E\{(\tilde{A}_c(\sigma)-A(\sigma))^2\}&=(l-W^Ts)^T(W^T\Sigma^{-1}W)^{-1}(l-W^Ts)+d'(\Omega-\Omega U'\Sigma^{-1} U\Omega)d\nn\\
&+(l'-s'W)\Delta_1\Sigma\Delta_1^T(l'-s'W)^T\nn\\
&+2(l'-s'W)\Delta_1\Sigma(s'+(l'-s'W)(W^T\Sigma^{-1}W)^{-1}W^T\Sigma^{-1})^T\nn\\
&-2(l'-s'W)\Delta_1 U\Omega d'+O(n^{-2})\label{g1g2g3}
\end{eqnarray}
where $s=d'\Omega^*U^T{\Sigma^*}^{-1}$.

Let $B=(B_1,\cdots,B_g)$ and $J=(J_1,\cdots,J_g)^T$ where
\bea
B_k^T&=&-(s'W-l')(W^T{\Sigma^*}^{-1}W+\Delta_\beta)^{-1}W^T{\Sigma^*}^{-1}\frac{\partial \Sigma^*}{\partial \sigma_k}{\Sigma^*}^{-1}W(W^T{\Sigma^*}^{-1}W+\Delta_\beta)^{-1}W{\Sigma^*}\nn\\
&&+(s'W-l')(W^T{\Sigma^*}^{-1}W+\Delta_\beta)^{-1}W^T{\Sigma^*}^{-1}\frac{\partial \Sigma^*}{\partial \sigma_k}{\Sigma^*}^{-1}\nn\\
&&-s'\frac{\partial \Sigma^*}{\partial \sigma_k}{\Sigma^*}^{-1}(I-W(W^T{\Sigma^*}^{-1}W+\Delta_\beta)^{-1}W{\Sigma^*}),\nn
\eea
and
$
J_k^T=\theta^T\Delta_\beta(W^T{\Sigma^*}^{-1}W+\Delta_\beta)^{-1}W^T{\Sigma^*}^{-1}\frac{\partial \Sigma^*}{\partial \sigma_k}+\theta^TWP\frac{\partial \Sigma^*}{\partial \sigma_k}P.
$
Further, denote $(\lambda_1,\cdots,\lambda_g)^T:=D^{-1}B$ and $G_i=P\frac{\partial \Sigma^*}{\partial \sigma_i} P$. The following theorem states a practical formula for the $ MSE\{\hat{A}_c(\hat{\sigma})\} $.

\btheorem
\label{thm3}
The MSE of prediction $\hat{A}_c(\hat{\sigma})$ for $A(\sigma)$ is
\begin{eqnarray*}
&MSE\{\hat{A}_c(\hat{\sigma})\}\nn\\
&\hspace{-3cm}=(l-W^Ts)^T(W^T\Sigma^{-1}W)^{-1}(l-W^Ts)+d'(\Omega-\Omega U'\Sigma^{-1} U\Omega)d +(l'-s'W)\Delta_1\Sigma\Delta_1^T(l'-s'W)^T\nn\\
&\hspace{-1.5cm}+2(l'-s'W)\Delta_1\Sigma(s'+(l'-s'W)(W^T\Sigma^{-1}W)^{-1}W^T\Sigma^{-1})^T -2(l'-s'W)\Delta_1 U\Omega d'+2tr\{(BD^{-1}J\Sigma)^2\}\nn\\
&\hspace{-2cm}+tr^2(\Sigma BD^{-1}J)
+tr(D^{-1}B^T\Sigma BD^{-1}\Sigma_w)+4\sum_{j=1}^g\sum_{l=1}^g\lambda_j^T\Sigma(G_j\Sigma G_j+G_l\Sigma G_j)\Sigma\lambda_l+o(n^{-1}),
\end{eqnarray*}
where $D=\left(\frac{\partial \mathcal{Q}_{VC,k}(\sigma)}{\partial\sigma_l}\right)_{kl}$, $\tilde{e}=(e_1,e_2,\cdots,e_g)^T$ and $e_k=Y^TP\frac{\partial \Sigma^*}{\partial \sigma_k}PY-tr(P\Sigma P\frac{\partial \Sigma^*}{\partial \sigma_k})$, and $\Sigma_w=(2tr(G_i\Sigma G_j \Sigma))_{i,j}$.
\etheorem

Therefore, an estimation of the $MSE\{\hat{A}_c(\hat{\sigma})\}$ can be obtained by plugging in the variance components estimate $\hat{\sigma}$ from (\ref{est:equa:variance}) into the MSE expression.

\subsection{Choice of smoothing parameters}

As mentioned earlier, the basic modeling framework considered in this article is similar to Chen and Wang (2011), however, there is differences in smoothing procedure of  fixed and random effects and in parameter estimation. We followed  their  smoothing parameter choices. For completeness we explained the procedure here briefly.

Chen and Wang (2011) proposed to estimate the variance component in $R$ by minimizing the following penalized log-likelihood, for given $\theta$ and $\alpha$,
\be
\ell_v(\sigma)=\log|R|+(Y-W\theta-U\alpha)^TR^{-1}(Y-W\theta-U\alpha)+\theta^T\Delta_\beta\theta+\alpha^T\Delta_{\nu}\alpha.
\ee

Decomposing each $W_k^T(t)=(W_{k(1)}^T(t),W_{k(2)}^T(t))^T$ and let $W_{k(1)}^T(t)$ be the first $r$ component of the $W_k^T(t)$ and $W_{k(2)}^T(t)$ be the rest $L_0$ component of $W_k^T(t)$.  Now collecting $W_{(1)}^T(t)=(W_{1(1)}^T(t),\cdots,W_{p(1)}^T(t))$ be a $pr$ components vector and $\theta_{(1)}$ be the corresponding coefficients. The smoothing parameter $\lambda=(\lambda_1,\cdots,\lambda_p)^T$ are chosen by minimizing the following marginal REML
\be
\label{eq:tuning_parameters}
\ell_m(\lambda)=\log|\Sigma_\lambda|+(Y-W_{(1)}\theta_{(1)})^T\Sigma^{-1}_\lambda(Y-W_{(1)}\theta_{(1)})+\log |W_{(1)}^T\Sigma^{-1}_\lambda W_{(1)}|
\ee
where $\Sigma_\lambda$ is the marginal covariance of $Y$, which is
$$
\Sigma_\lambda=R+U\Omega U+V_{(2)}
$$
where $V_{(2)}=\sum_{k=1}^pV_{k(2)}$, $V_{k(2)}=\lambda_{k}^{-1}W_{k(2)}^TW_{k(2)}$ and $W_{k(2)}=(W_{k(2)}(t_{11}),\cdots, W_{k(2)}(t_{nm}))^T$.

The smoothing parameters $\eta=(\eta_1,\cdots,\eta_q)^T$ are chosen by minimizing the following marginal log-likelihood, for given $\alpha=\hat{\alpha}$
\be
\ell_m(\eta)=\log|R|+(Y-W\theta-U\alpha)^TR^{-1}(Y-W\theta-U\alpha)+\alpha^T\Delta_{\nu}\alpha+\log|H|-\log|\Delta_\nu|. \nn
\ee
where
$$
H=\frac{1}{2}\frac{\partial ^2\ell_v(\sigma)}{\partial\alpha\partial\alpha^T}=U^TR^{-1}U+\Delta_{\nu}
$$
for any symmetric matrix $\Delta_{\nu}$. Because $|\Delta_\nu|=\prod_{i=1}^n\prod_{k=1}^q\eta_k^{L}|\Delta_{\nu ik}|$, taking derivative of $\ell_m(\eta)$ with respect to $\eta_k$, gives
$$
\sum_{i=1}^n\alpha_{ik}^T\Delta_{\nu ik}\alpha_{ik}+tr(H^{-1}\tilde{\Delta}_{\nu}^{(k)})-\frac{L}{\eta_k}=0
$$
where $\tilde{\Delta}_{\nu}^{(k)}=\diag\{\tilde{\Delta}_{\nu 1}^{(k)},\cdots,\tilde{\Delta}_{\nu n}^{(k)}\}$ and $\tilde{\Delta}_{\nu i}^{(k)}=\diag\{0,0,\cdots, \Delta_{\nu ik},\cdots, 0\}$ for $i=1,\cdots,n$.
Then
$$
\hat{\eta}_k=L\,{\left(\sum_{i=1}^n\alpha_{ik}^T\Delta_{\nu ik}\alpha_{ik}+tr(H^{-1}\tilde{\Delta}_{\nu}^{(k)})\right)}^{-1}.
$$

\section{Numerical Findings}\label{numerical study}
We investigated the finite sample performance of the proposed method through simulation and real data examples. The simulation study is also designed to verify the asymptotic behavior of the approximated prediction error formula and and related coverage errors.
\subsection{Simulation Study}
We considered the following mixed model setup
\be
\label{equ3.1}
Y(t_{ij})=X(t_{ij})\beta(t_{ij})+Z(t_{ij})\nu(t_{ij})+\epsilon(t_{ij})\quad\mbox{$i=1,\cdots,n$ and $j=1,\cdots,m_i$}
\ee
where $m_i=1$ for all $i=1, . . ., n$. Three different values of  $n$ was chosen, 50, 100 and 200. We generated the
 time points $t_{ij}$ independently from Uniform(0,1). Let $X(t_{ij})=1+0.5t_{ij}+e_{ij}, Z(t_{ij})= (0.1)\,\left( -(0.8415/2) + \sin(t_{ij})+u_{ij} \right), \beta(t_{ij})=2\,\cos(t_{ij})$.
 Set $e_{ij}\sim N(0,0.5^2)$ and $u_{ij}\sim N(0,0.4^2)$.
 We simulated $\nu(t)$ from a Gaussian process with mean 0 and covariance $\cov(\nu(t_i),\nu(t_j))= B^T_{t_i}\,\rho^{|i - j|}\, B_{t_j}$ where $\rho=0.4$, and $\epsilon(t_{ij})$ from a Gaussian process with mean zero and $\cov(\epsilon(t),\epsilon(s))=\sigma_{\epsilon}^2\,\delta_{st} $, where $ \delta_{st} = 1 $ if $ s=t $, and $ 0 $ otherwise. We set $\sigma_{\epsilon}=1$.   We call this set up as Case I. As a second case, we kept everything same except the covariance structure of the random effects. Specifically, we took $\cov(\nu(s),\nu(t))= \rho^{|s - t|}$ with $ \rho = 0.4 $. We call this as Case II. We also examined the performance for more fluctuated mean function where $\beta(t_{ij})= \cos(2\pi t_{ij})$. We call this as case III.

The main purpose of this simulation study is to evaluate the performance of mean square error of the predictor of mixed effects that measuring subject (curve) specific means. For example, we considered $A_i=\bar{X}_i\beta(t_0)+\bar{Z}_i\nu_i(t_0), i=1,\cdots,n$ where $t_0$ is one of the time points $ t_{ij} $ (for example $ t_{11} $), and $\bar{X}_i$ and $\bar{Z}_i$ are means of $X_{ij}$ and $Z_{ij}$ respectively. In particular, here $p=1$ and $q=1$, $l_0=\bar{X}_i$ and $d_{i0}=\bar{Z}_i$ and rest of $d_{j0}=0$ for $j\neq i$. Similar to Chen and Wang (2007), we used the following algorithm to estimate $\beta,\alpha,\Omega$ and $\sigma$. We fixed the tuning parameters $\lambda$ and $\eta$ and set an initial value of $\Omega_{(0)}=\diag\{1,\cdots,1\}$, $\sigma^2_{\epsilon(0)}=1$, $\Omega^*_{(0)}=(\Omega_{(0)}^{-1}+\Delta_{\nu})^{-1}$ and $\Sigma_{i(0)}^*=U_i(\Omega^{-1}_{(0)}+\Delta_{\nu i})^{-1}U_i^T+\sigma^2_{\epsilon}I$. Then the initial estimates of $\alpha$ and $\theta$  are $$\tilde{\alpha}_{(0)}=\Omega^*_{(0)}U^T{\Sigma^*}^{-1}_{(0)}(Y-W\tilde{\theta}_{(0)})\;\;\mbox{and}\;\; \tilde{\theta}_{(0)}=(\sum_{i=1}^n W_i^T\Sigma_{i(0)}^* W_i+\Delta_\beta)^{-1}\sum_{i=1}^n W_i^T{\Sigma_i^*}^{-1}_{(0)}Y_i.$$ respectively.
We then repeat the following steps until all the parameter estimates converge
\begin{itemize}
\item[] {\bf Step 1:} estimate $\sigma^2_{\epsilon}$ through the estimating equation given in (\ref{est:equa:variance}).
\item[] {\bf Step 2:} estimate $\tilde{\theta}$ and $\tilde{\alpha}$ through (\ref{theta-esti}) and (\ref{alpha-esti}), and $\hat{\Omega}$ by
$$
\hat{\Omega}=\frac{1}{n}\sum_{i=1}^n\left\{\tilde{\alpha}_i\tilde{\alpha}_i^T+\hat{\Omega}^*-\hat{\Omega}^*U_i^TM_iU_i\hat{\Omega}^*\right\}
$$
where $M_i=\hat{\Sigma}_i^{-1}-\hat{\Sigma}_i^{-1}W_i(\sum_{i=1}^nW_i^T\hat{\Sigma}_i^{-1}W_i+\Delta_\beta)^{-1}W_i\hat{\Sigma}_i^{-1}$.
\end{itemize}

Plugging in the estimates of $\theta,\alpha,\sigma_{\epsilon}$ and $\Omega$ from the previous algorithm,  obtain a biased corrected estimation of $\hat{A}$ by $\tilde{A}_c(\sigma)$ given in (\ref{bias-corr-prediction}).

We then compute the following three quantities:

\begin{itemize}
\item[]
The true MSE: computed by $\frac{1}{K}\sum_{k=1}^K(\hat{A}_i^{(k)}-A_i^{(k)})^2, i=1,\cdots, n$ where $K$ is the number of replicated data sets and $\hat{A}_i^{(k)}$ is the prediction of $A_i^{(k)}$ in the $k$-th replicate. In all cases we took $K=600$ replication.

%\item[]
%Estimate MSE with true variance component:  computed by the formula given in (\ref{g1g2g3}) by taking $\sigma$ to be the true value but with estimated $\Omega$. [THIS IS NOT REPORTED, RIGHT? Yes, This is not reported here. This part was actually just for us to check.]

\item[]
Estimate MSE with estimated variance components:  computed by the formula given in (\ref{g1g2g3}) using estimated $\sigma$ and $\Omega$.
\end{itemize}

Along with the  mean square prediction errors, we calculated 95\% prediction coverage error where the prediction interval was calculated as  $\widehat{A} \, \pm \, 2\,{( \mbox{estimated} \,\, \mbox{MSE})}^{1/2}$.

The reported values in the table \ref{table:caseI,II,III} are the averages of the prediction coverage over all the individuals.
%  We also computed confidence coverage for the simultaneous confidence band  based on the method stated in Xue and Zhu (2007) \cite{Xue_Zhu_2007}, section 4.3 for different number of time points ($ M $). We took $ M=6$, equally spaced in $ [0, 1]$.
  The relative bias is the relative difference between the true MSE and the estimated MSE,  averaged over  replications and  then averaged over the individuals. Numbers in the bracket are the standard errors averaged over all the subjects.

\begin{sidewaystable}
%\begin{table}[ht]
\caption{Output under the proposed method. The computation time for cases I, II and III are 37498, 52698 and 42659 sec respectively. } % title of Table
\centering % used for centering table
\begin{tabular}{c c c c c c} % centered columns (4 columns)
\hline\hline %inserts double horizontal lines
n & $ \wh{\sigma_{\epsilon}} $ & Prediction Coverage & Relative Bias & True MSE \\ [0.5ex] % inserts table
%heading
\hline % inserts single horizontal line
Case I: & & & & & \\ [1ex]
50 & 0.9915(0.1034) & 0.9454(0.0044)  & 0.0509(0.0325) & 0.1015(0.0215)   \\ % inserting body of the table
100 & 0.9936(0.0732) & 0.9488(0.0051) & 0.0749(0.0365) & 0.0599(0.0126) \\
200 & 0.9968(0.0521) & 0.9558(0.0062) & 0.0460(0.0332) & 0.0222(0.0048) \\ [1ex] % [1ex] adds vertical space
Case II: & & & & & \\ [1ex]
50 & 0.9918(0.1043) & 0.9431(0.0058)  & 0.0560(0.0342) & 0.1024(0.0217)   \\ % inserting body of the table
100 & 0.9937(0.0733) & 0.9453(0.0059) & 0.0859(0.0385) & 0.0608(0.0129) \\
200 & 0.9971(0.0521) & 0.9503(0.0067) & 0.0383(0.0312) & 0.0229(0.0050) \\ [1ex] % [1ex] adds vertical space
Case III: & & & & & \\ [1ex]
50 & 0.9919(0.1042) & 0.9426(0.0067)  & 0.0599(0.0356) & 0.1029(0.0218)   \\ % inserting body of the table
100 & 0.9941(0.0734) & 0.9575(0.0057) & 0.0969(0.0392) & 0.0616(0.0131) \\
200 & 0.9974(0.0521) & 0.9660(0.0064) & 0.0431(0.0345) & 0.0236(0.0051) \\ [1ex] % [1ex] adds vertical space
\hline %inserts single line
\end{tabular}
\label{table:caseI,II,III} % is used to refer this table in the text
%\end{table}
\end{sidewaystable}

In summarizing the tables, the performance of prediction error estimation is very satisfactory. The relative bias in prediction error is less than 10\%. Both the prediction coverage and confidence coverage are fairly close to the nominal level.

\subsection{Comparison with Guo  (2002a)}

Guo (2002a) reported prediction intervals based on Wahba (1983).  In this section, we compare the numerical performance of the proposed method with Guo (2002a) in terms of subject specific prediction intervals. For this purpose, we considered the same model (\ref{equ3.1})
%$$
%Y(t_{ij})=X(t_{ij})\beta(t_{ij})+Z(t_{ij})\nu(t_{ij})+\epsilon(t_{ij})\quad\mbox{$i=1,\cdots,n$ and $j=1,\cdots,m_i$}
%$$
with $m_i=6$ for all $i=1, . . ., n$, and $n$ was chosen as 25 and 50.
 %We simulate independent and identically distributed (IID) time points $t_{ij}$ from Uniform(0,1).
  We took, $X(t_{ij}) = (4.5)e^{i/n}, Z(t_{ij}) = (0.1)e^{-i/n}, \beta(t_{ij}) = \cos(2 \pi t_{ij})$.
%  [Why did you change X here ? Since the sample size is
%  small here, $ n=25, 50 $, signal-to-noise ratio must be adjusted, so I had to increase the magnitude of the design matrix $ X $. So, I chose these mean functions. It is also good that we are checking our simulations with different types of mean functions, I guess.]
 %We simulated $\nu(t)$ from a Gaussian process with mean 0 and covariance $\cov(\nu(t_i),\nu(t_j))= \rho^{|t_i - t_j|}$ where $\rho=0.4$, and $\epsilon(t_{ij})$ from a Gaussian process with mean zero and $\cov(\epsilon(t),\epsilon(s))=\sigma_{\epsilon}^2\,\delta_{st} $, where $ \delta_{st} = 1 $ if $ s=t $, and $ 0 $ otherwise, and
We set $\sigma_{\epsilon}=0.14$ for $n=25$ but changed to $0.12$  for $n=50$ to keep the signal-to-noise ratio comparable.
Then we predicted the response at the time points $t_{ij}$ and calculated coverage and length of the prediction intervals under both the methods. For Guo (2002a), we used the SAS code provided by Liu and Guo (2011). Since the SAS code takes considerable longer time compared to our  MATLAB code, we used only $K=100$ replicates in this comparison. The table \eqref{table:sim_comparison} reported numerical values that were averaged over all individuals.

The 3rd and 4-th column correspond to the proposed method where as the 5-th and 6-th columns are correspond to Guo (2002a). The proposed method clearly outperform both the coverage and length of the prediction intervals. The coverage under proposed method is closed to the nominal level whereas that is about only 50\% under Guo (2002a). The length of prediction interval under Guo (2002a) is about twice compared to the proposed method. However, in terms of prediction bias, both the methods are nicely comparable. Although the performance of the proposed method is remarkable in terms of prediction error, we like to re-iterate Guo (2002a)'s original development was not meant for prediction error, rather model estimation. Thus the result is not completely unexpected. Furthermore, the numerical study indicates the importance of the development of uncertainty measures under the frequentist approach.
Since the numerical difference is remarkable, we explain below how the quantities are calculated, for clarification.

%[So I added the relative prediction bias to the Table (2) below. They are almost identical to Guo2002, which was expected by Professor Maiti. Case when $ n=50 $ for Guo2002 takes much more time. I will add the results for this case also as soon as runs are finished.]
%
%\textbf{Explanation of how the quantities in the table were calculated (For simulations):}

For fixed time point $ t $, we wish to predict the quantity $ A_{ij} $ for individual $ i $ in replication $ j $. Note that the number of individuals and replications here are denoted by $ n $ and $ K $, respectively. Denote the predicted value by $ \widehat{A_{ij}} $. Then we derived the prediction intervals under each methods as described above. Denote this intervals by $ I_{ij} $. Then for a time point $t$, we calculated the following:
\begin{eqnarray*}
\mbox{Prediction Coverage} (t) &=& \frac{1}{nK} \sum_{i=1}^{n} \sum_{j=1}^{K} \delta_{I_{ij}}(A_{ij})\\
\mbox{Length}(t) &=& \frac{1}{nK} \sum_{i=1}^{n} \sum_{j=1}^{K} \mbox{length}(I_{ij})\\
\mbox{Prediction Bias}(t) &=& \frac{1}{nK} \sum_{i=1}^{n} \sum_{j=1}^{K} \left\vert \dfrac{A_{ij} - \widehat{A_{ij}}}{A_{ij}}\right\vert
\end{eqnarray*}

where $ \delta_A(b)=1 $ if $ b \in A $, and $ 0 $ otherwise.

\begin{sidewaystable}
%\begin{table}[ht]
\caption{Comparison of the proposed method with Guo (2002a) in terms of prediction intervals} % title of Table
\centering % used for centering table
\begin{tabular}{c c c c c c c c} % centered columns (4 columns)
\hline\hline %inserts double horizontal lines
n & time & Pred Cov & Length & PredBias & PredCov(Guo2002a) & Length(Guo2002a) & PredBias(Guo02a) \\ [0.5ex] % inserts table
%heading
\hline % inserts single horizontal line
25  & 0.9420 & 0.9000  & 0.0850 & 0.0152  & 0.4919  & 0.1677 & 0.0150 \\
    & 0.4491 & 0.9464  & 0.0893 & 0.0151  & 0.4238  & 0.1495 & 0.0153 \\
    & 0.5752 & 0.9536  & 0.0912 & 0.0164  & 0.4119  & 0.1491 & 0.0166 \\
    & 0.0965 & 0.9572  & 0.0871 & 0.0171  & 0.507   & 0.1837 & 0.0173 \\
    & 0.9437 & 0.9124  & 0.0865 & 0.0150  & 0.5065  & 0.1684 & 0.0150 \\
    & 0.7573  & 0.9664  & 0.0911 & 0.8821   & 0.4508  & 0.1520 & 1.0050 \\

50  & 0.9420 & 0.9706  & 0.0705 & 0.0130  & 0.4624  & 0.138 & 0.0269  \\
    & 0.4491 & 0.9740  & 0.0675 & 0.0124  & 0.38    & 0.112 & 0.0269  \\
    & 0.5752 & 0.9780  & 0.0769 & 0.0134  & 0.38    & 0.111 & 0.0279  \\
    & 0.0965 & 0.972   & 0.0688 & 0.0144   & 0.48    & 0.146 & 0.0284  \\
    & 0.9437 & 0.9680  & 0.0714 & 0.0129  & 0.45    & 0.138 & 0.0265  \\
    & 0.7573  & 0.9710  & 0.0693 & 0.7034    & 0.39    & 0.115 & 0.8857  \\ [1ex] % [1ex] adds vertical space

\hline %inserts single line
\end{tabular}
\label{table:sim_comparison} % is used to refer this table in the text
%\end{table}
\end{sidewaystable}

\subsection{Real Data Examples}

In this section, we apply the proposed method to two real data sets for illustration.
\vspace*{2ex}

\noindent
{\it Example 1.} We considered the case study of lung function (FEV1) from a longitudinal epidemiologic study (Fitzmaurice, Laird and Ware, 2012). 1 to 7 seven repeated measurements on FEV1 were taken on each of the 133 (sampled) subjects aged 36 or older. Fitzmaurice {\it et al.} (2012) argued for a cubic polynomial mean model for regressing FEV1 on smoking behaviors. An appropriateness of a  varying coefficient model can be tested by comparing the following two models.
\begin{eqnarray*}
H_0 &:& Y(t) = X(t)\theta_0 + Z(t)v(t) + \epsilon(t)\\
H_1 &:& Y(t) = X(t)\theta_0 + \sum_{l=1}^{L}X(t)B_l(t)\theta_l + Z(t)v(t) + \epsilon(t)
\end{eqnarray*}
This equivalently testing the hypothesis
\begin{equation*}
H_0 : \theta_1 = ... = \theta_L=0
\end{equation*}

The test statistics here is
\begin{equation*}
T_n = 2 \, \log \dfrac{\mbox{likelihood } H_1}{\mbox{likelihood } H_0}
\end{equation*}
This is asymptotically a Chi squared random variable with $ L $ degrees of freedom under $ H_0 $. For our case $ T_n = 700.5114 - 684.0968 = 16.4146 $, which gives the p--value $ 0.0116934 $ when $ L=6 $. This justifies a functional model such as  (\ref{equ3.1}).

\begin{figure}[ht]
\centering
\vskip-30pt
%\rotatebox{270}
{\includegraphics[scale=0.8]{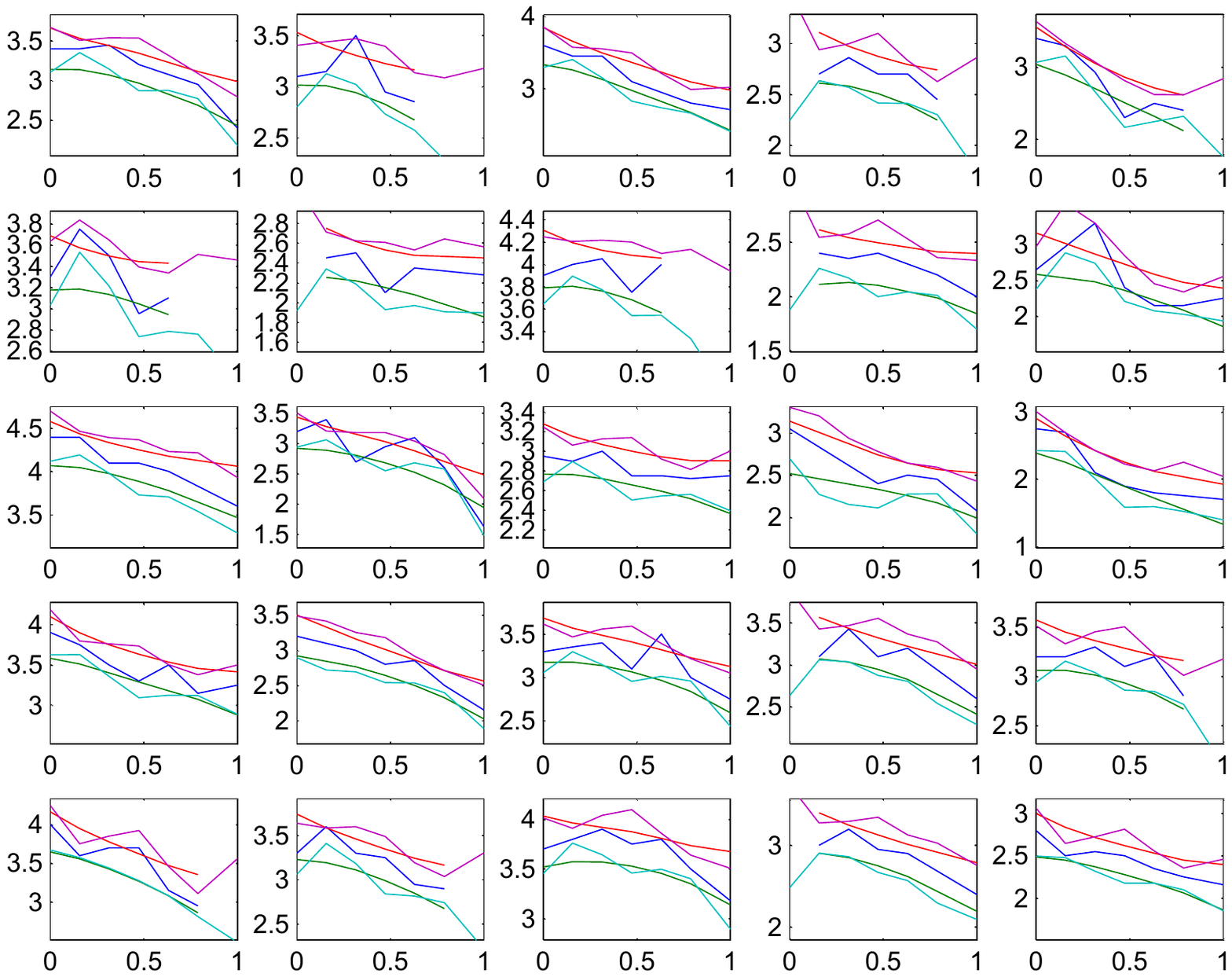}}
\vspace{-6.5cm}
%\vskip-30pt
\caption{FEV1 data: red and green: prediction band by Guo2002a. purple and light blue: proposed prediction band}
\label{fig:FEV1 comparison}
\end{figure}

The Figure \eqref{fig:FEV1 comparison} presented the predicted curves and their 95\% prediction intervals for the first 25 individuals.
%[Also, I added the overlaid plot for comparing with Guo2002 in figure \eqref{fig:FEV1 comparison}].
 The fit seems reasonable. The table \eqref{Table:Coverage_FEV1} reported the prediction coverage of the proposed method compared to Guo (2002a) in all the time points.  For a fixed time point, the prediction coverages were calculated by checking the proportion of  subject specific intervals cover the true value.

\begin{sidewaystable}
%\begin{table}[ht]
\caption{Prediction coverage and length for FEV1 data } % title of Table
\centering % used for centering table
\begin{tabular}{c c c c c c c} % centered columns (4 columns)
\hline\hline %inserts double horizontal lines
time points & PredCov & PredCov(Guo) & Length & Length(Guo) & PredBias & PredBias(Guo)\\ [0.5ex] % inserts table
%heading
\hline % inserts single horizontal line
1 & 1 & 0.9167 & 0.7033 & 0.5242 & 0.0004 & 0.0361\\
2 & 0.8115 & 0.7705 & 0.2902 & 0.4183 & 0.0001 & 0.0411\\
3 & 0.8974 & 0.7692 & 0.4480 & 0.3730 & 0.0003 & 0.0418\\
4 & 0.9739 & 0.8000 & 0.6778 & 0.3700 & 0.0001 & 0.0496\\
5 & 0.9273 & 0.8455 & 0.4547 & 0.3892 & 0 & 0.0407\\
6 & 0.9278 & 0.8969 & 0.4210 & 0.4200 & 0.0001 & 0.0404\\
7 & 1 & 0.9510 & 0.7400 & 0.5559 & 0.0009 & 0.0395\\
averages & 0.934 & 0.85 & 0.5335 & 0.4358 & 0.0003 & 0.0413\\[1ex]
\hline %inserts single line
\end{tabular}
\label{Table:Coverage_FEV1} % is used to refer this table in the text
%\end{table}
\end{sidewaystable}

%
%\begin{figure}[ht]
%\centering
%\vskip-30pt
%%\rotatebox{270}
%{\includegraphics[scale=0.8]{prediction_bands.pdf}}
%\vskip-30pt
%\caption{FEV1 data combined with the prediction bands for the first 25 subjects.}
%\label{fig:FEV1 prediction}
%\end{figure}

\vspace*{2ex}

\begin{figure}[ht]
\centering
\vskip-30pt
%\rotatebox{270}
{\includegraphics[scale = 0.8]{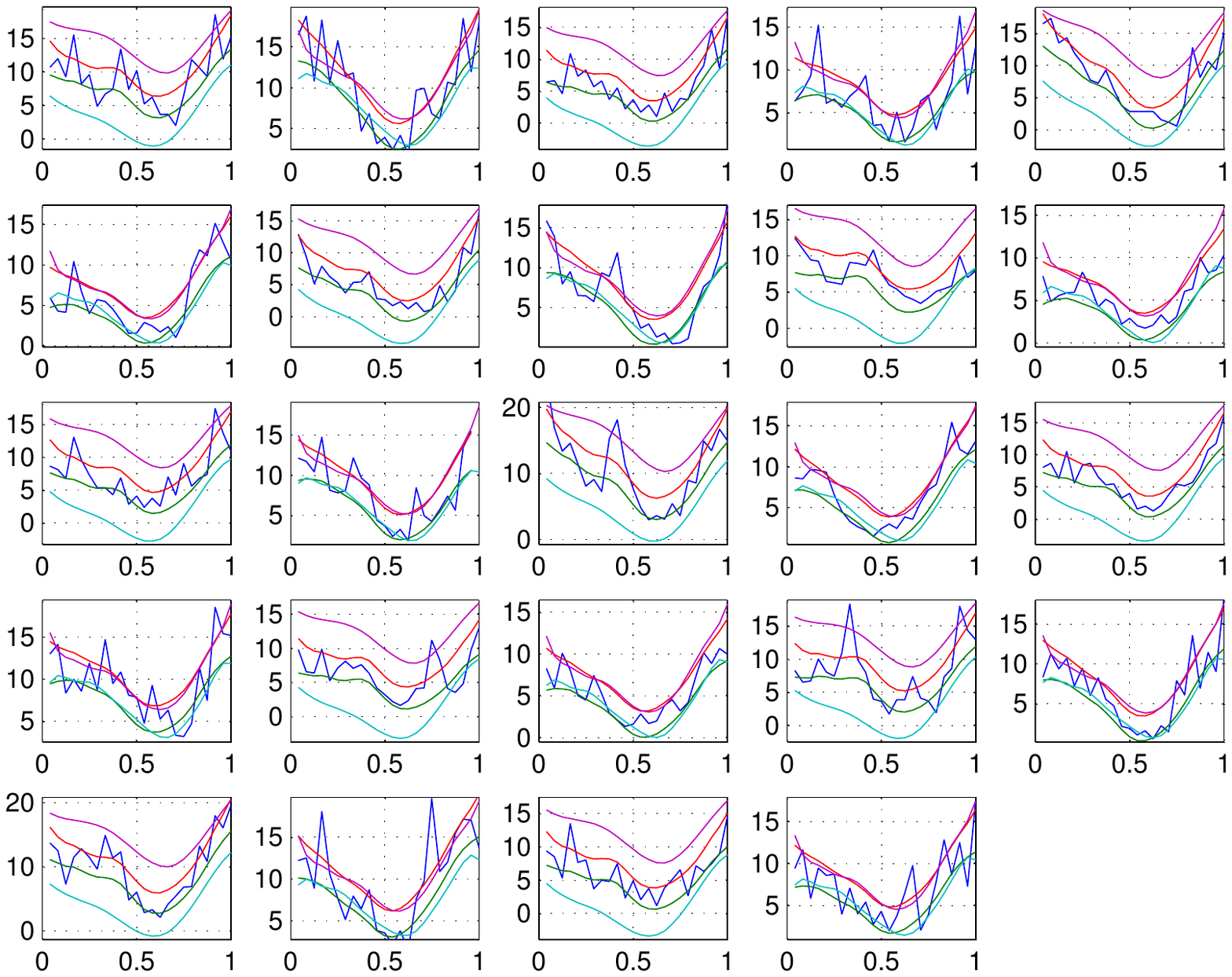}}
\vspace{-6.5cm}
%\vskip-30pt
\caption{Cortisol data: red and green: prediction band by Guo2002a. purple and light blue: proposed prediction band}
\label{fig:Cortisol comparison}
\end{figure}

\noindent
{\it Example 2. }
 We consider another data example of modeling  blood concentrations of cortisol as considered by  (Guo, 2002a). There were 22 patients, 11 with fibromyalgia (FM) and 11 normal (the plot in Guo (2002a) showed 24 patients, 12 on each group)  and their blood samples  were observed every hours over a 24 hours period of time. The objective is to model concentration of cortisol on FM patients. Guo (2002a) argued for a functional model for this data set. We fit our model and came to the similar conclusion that the FM group has significantly higher cortisol.
%   .... [Produce plots of $\beta(t)$ for the two groups along their confidence intervals, or some kind of support of this claim: Figures \eqref{fig:Cortisol_splitplot} and \eqref{fig:Cortisol_splitplot_withbands} are plotted for this part.].
 Figure \eqref{fig:Cortisol comparison} shows the subject specific prediction along with the prediction intervals.

%[PL. overlay Guo on the same plot: I added the overlaid plot for comparing with Guo2002 in figure \eqref{fig:Cortisol comparison}]. [give the explanation how did you calculate the \eqref{Table:Coverage_Cortisol}:It's exactly the same as Table (3).]
%
%{\color{red} Like the previous example, also report the prediction bias here: Done! Actually I put the relative prediction bias in both cases.}

\begin{figure}[ht]
\centering
\vskip-30pt
%\rotatebox{270}
{\includegraphics[scale=0.8]{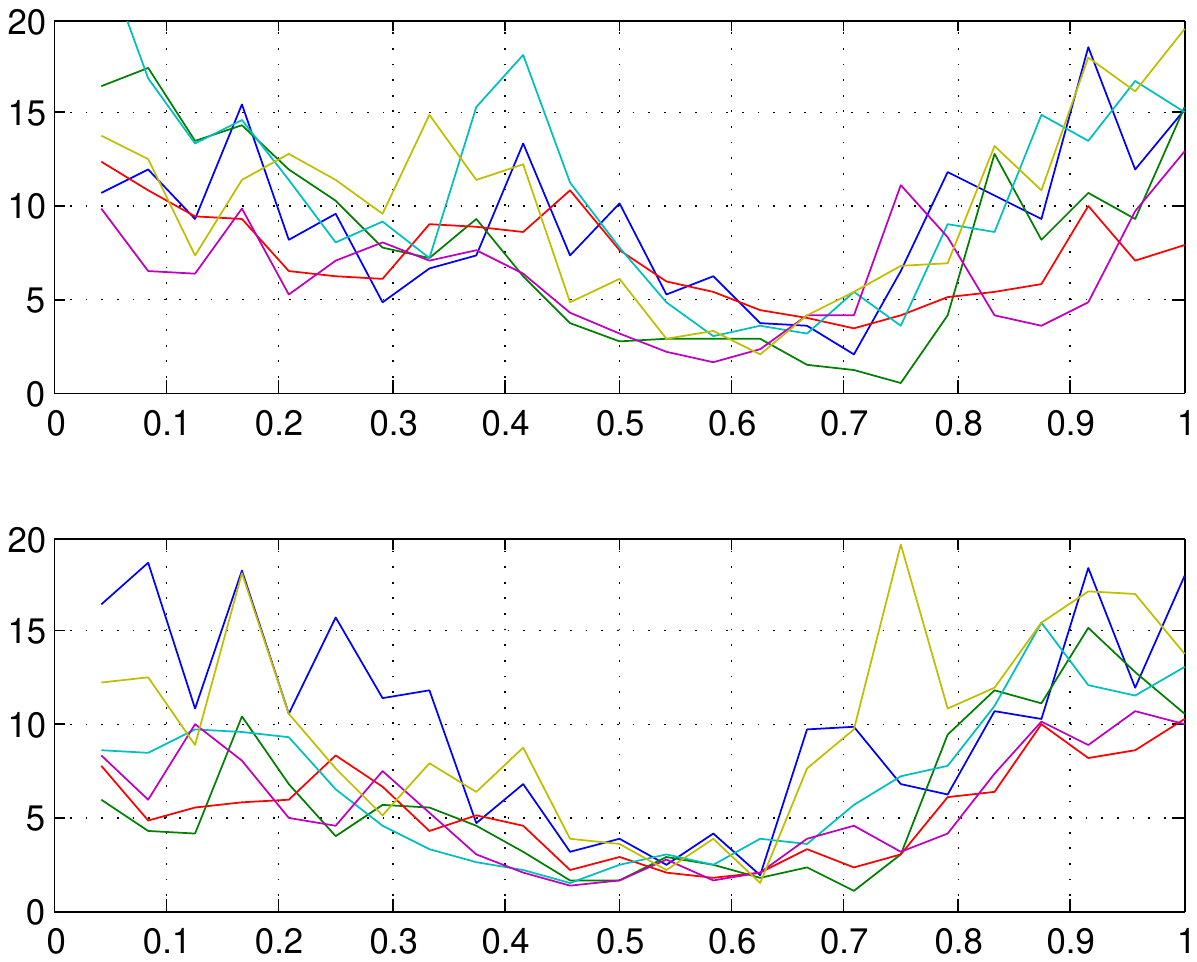}}
\vspace{-7.5cm}
%\vskip-30pt
\caption{Cortisol data: top: patient group. down: control group. 12 graphs for each.}
\label{fig:Cortisol_splitplot}
\end{figure}

\begin{figure}[ht]
\centering
\vskip-30pt
%\rotatebox{270}
{\includegraphics[scale=0.8]{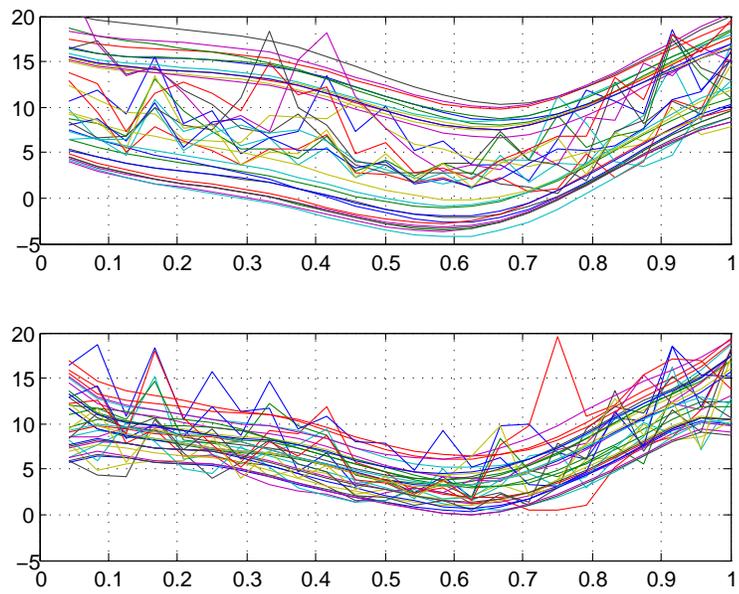}}
\vspace{-7.5cm}
%\vskip-30pt
\caption{Cortisol data with prediction bands: top: patient group. down: control group.}
\label{fig:Cortisol_splitplot_withbands}
\end{figure}

%Similar to the first example, we use the following model:
%\begin{align}
%\label{FMEM}
%Y(ij)=X(ij)\beta(t_{ij})+\nu_i(t_{ij})+{\epsilon}_{ij}, \hspace{2em} {\epsilon}_{ij} \sim N(0, {\sigma}^2_{\epsilon})
%\end{align}
%where  $ Y(ij)\, (i=1, ..., 24,\, j= 1, ..., 24)$ is the Cortisol concentration of the $ i^{th} $ subject at the $ j^{th} $ hour; $ X(ij)=1 $ if the $ i^{th} $ subject is in the patient group, and  $ X(ij)=0 $ if he/she was in the control group for all $ j $'s. Both $ \beta(t_{ij}) $ and $ \nu_i(t_{ij}) $ are modeled using the B-spline basis functions, as discussed in section \eqref{Preliminary}. Here we are interested in finding prediction for
%\begin{align}
%A_{ij}= X(ij) \beta(t_{ij})+\nu_i(t_{ij})
%\end{align}
%for different $ i $'s and $ j $'s, which is practically the mean structure of the model. Figure \eqref{fig:Cortisol prediction} and table \eqref{Table:Coverage_Cortisol} show the plot of the data together with the prediction bands for all the 24 patients, and the prediction coverage for all the time points, respectively. Also, in table \eqref{Table:Coverage_Cortisol} we compare the $ 95\% $ confidence intervals proposed in (Liu and Guo, 2011 \cite{Liu_Guo_2011}) with the prediction bands calculated using the method discussed in this article.

\begin{sidewaystable}
%\begin{table}[ht]
\caption{prediction coverage for Cortisol data } % title of Table
\centering % used for centering table
\begin{tabular}{c c c c c c c} % centered columns (4 columns)
\hline\hline %inserts double horizontal lines
t & PredCov & PredCov(Guo2002a) & Length & Length(Guo2002a) & PredBias & PredBias(Guo2002a)\\ [0.5ex] % inserts table
%heading
\hline % inserts single horizontal line
1 & 0.9167 & 0.625 & 8.4782 & 4.9302 & 0.0369 & 0.1469\\
2 & 0.8333 & 0.75 & 7.1589 & 4.0079 & 0.0372 & 0.1729\\
3 & 0.7500 & 0.5417 & 7.0751 & 3.5531 & 0.0358 & 0.2998\\
4 & 0.7500 & 0.5 & 7.2737 & 3.2912 & 0.0381 & 0.1999\\
5 & 0.9167 & 0.625 & 7.2601 & 3.1525 & 0.0347 & 0.1960\\
6 & 0.8750 & 0.7083 & 7.1888 & 3.0972 & 0.0356 & 0.1976\\
7 & 0.8750 & 0.5417 & 7.3709 & 3.0870 & 0.0294 & 0.3102\\
8 & 0.8750 & 0.6667 & 7.7112 & 3.0953 & 0.0327 & 0.2156\\
9 & 0.9167 & 0.6667 & 7.8958 & 3.1079 & 0.0335 & 0.2630\\
10 & 0.8750 & 0.6667 & 7.8527 & 3.1186 & 0.0371 & 0.2315\\
11 & 0.8750 & 0.4583 & 7.6627 & 3.1258 & 0.0334 & 0.5396\\
12 & 0.9583 & 0.875 & 7.4458 & 3.1293 & 0.0357 & 0.2083\\
13 & 0.8750 & 0.7083 & 7.2942 & 3.1293 & 0.0300 & 0.4598\\
14 & 0.9167 & 0.8333 & 7.1945 & 3.1257 & 0.0319 & 0.2549\\
15 & 0.8750 & 0.75 & 7.0657 & 3.1185 & 0.0251 & 0.6692\\
16 & 0.8750 & 0.7083 & 6.8582 & 3.1079 & 0.0253 & 0.3263\\
17 & 0.7083 & 0.7083 & 6.6018 & 3.0956 & 0.0120 & 0.8806\\
18 & 0.8333 & 0.5417 & 6.4041 & 3.0880 & 0.0326 & 0.9020\\
19 & 0.8333 & 0.625 & 6.3144 & 3.0996 & 0.0368 & 0.5168\\
20 & 0.8750 & 0.5833 & 6.2544 & 3.1571 & 0.0362 & 0.2335\\
21 & 0.8333 & 0.7083 & 6.1151 & 3.2987 & 0.0355 & 0.2823\\
22 & 0.7917 & 0.4167 & 5.9236 & 3.5648 & 0.0386 & 0.2051\\
23 & 0.8333 & 0.625 & 6.1485 & 4.0283 & 0.0369 & 0.2029\\
24 & 0.9130 & 0.4783 & 7.9552 & 4.9665 & 0.0355 & 0.1972\\
averages & 0.8575 & 0.6380 & 7.104 & 3.395 & 0.03319 & 0.3380\\[1ex]
\hline %inserts single line
\end{tabular}
\label{Table:Coverage_Cortisol} % is used to refer this table in the text
%\end{table}
\end{sidewaystable}

%\begin{figure}[ht]
%\centering
%\vskip-30pt
%%\rotatebox{270}
%{\includegraphics[scale=0.8]{Cortisol_prediction_new.pdf}}
%\vskip-30pt
%\caption{Cortisol data combined with the prediction bands for all the 24 patients.}
%\label{fig:Cortisol prediction}
%\end{figure}

%

\section{Discussion}\label{Discussion}

The main motivation of this paper is to develop computationally simple yet theoretically valid uncertainty estimation in the context of individualized prediction using functional data. Although our approach is bias-corrected plug-in method, there could be additional bias due to plug-in the estimated parameters. It is possible to achieve further accuracy, however, that might introduce additional variability on the estimated uncertainty, thus the resulting formulas may loose their practicality. Nevertheless, the corrections we develop here are highly satisfactory as the maximum relative bias is less than 10\%. The theoretically valid calculation of prediction intervals could be a dunting task in this context. However, if we simply apply the naive $\pm$ two standard type formula where the standard error formula is theoretically validated, the empirical coverage probabilities and the prediction lengths are much more improved than previously reported values.
 
The proposed method is implemented in MATLAB. Most of the computations of the proposed method can be handled easily due to the closed form solutions derived here. However, there are two parts which need numerical approximations, and high percentage of computation time in the method is due to these two optimization components. The first one is estimating the variance components $ \sigma $ in the $ Q $ function by finding its root as mentioned in the equation (\ref{est:equa:variance}). This part is programmed in MATLAB using the function ``fzero". The second part is to estimate the tuning parameters $ \lambda $ as the minimizer of the function $ \l_m(\lambda) $ stated in the equation (\ref{eq:tuning_parameters}). The optimization function ``fminbnd" in MATLAB is utilized to perform this numerical approximation. In contrast, Liu and Guo (2011) have developed a SAS code to implement their method. The computation time for their developed macro ``fmixed" is considerably higher than our MATLAB code. This might be due to the differences between the optimization method used in our code and the ones utilized in ``PROC MIXED".

\appendix

%  To get the journal style of heading for an appendix, mimic the following.

\section{}
\subsection{Proofs}

In this Appendix, we provide technical proofs to Theorems in the paper.
%\bigskip

\noindent {\bf Proof of Theorem \ref{thm1}:} By definition,
$\tilde{\theta}=(\sum_{i=1}^n W_i^T\Sigma_i^* W_i+\Delta_\beta)^{-1}\sum_{i=1}^n W_i^T{\Sigma_i^*}^{-1}Y_i$ and $\tilde{\alpha}=\Omega^*U^T{\Sigma^*}^{-1}(Y-W\tilde{\theta})$
Then
\begin{align*}
\tilde{\theta}-\theta&=-(\sum_{i=1}^n W_i^T\Sigma_i^* W_i+\Delta_\beta)^{-1}\Delta_\beta W_i\theta+(\sum_{i=1}^n W_i^T\Sigma_i^* W_i+\Delta_\beta)^{-1}\sum_{i=1}^n W_i^T{\Sigma_i^*}^{-1}(U_i\alpha_i+e_i)
\end{align*}
and
\begin{align*}
\tilde{A}(\sigma)-(l^T\theta+d^T\alpha)&=-l'(\sum_{i=1}^n W_i^T\Sigma_i^* W_i+\Delta_\beta)^{-1}\Delta_{\beta}\theta\\
&\quad+l'(\sum_{i=1}^n W_i^T\Sigma_i^* W_i+\Delta_\beta)^{-1}\sum_{i=1}^n W_i^T{\Sigma_i^*}^{-1}(U_i\alpha_i+e_i)\\
&\quad+d'\Omega^*U^T{\Sigma^*}^{-1}(Y-W{\theta})+d'\Omega^*U^T{\Sigma^*}^{-1}W(\theta-\tilde{\theta})-d^T\alpha.
\end{align*}
After taking expectation,  it is easy to see the conclusion.
%\bigskip

\noindent {\bf Proof of Theorem \ref{thm2}:} From the derivation in Theorem \ref{thm1}, it is easy to see that the bias of $\tilde{A}_c(\sigma)$ is
\begin{align*}
E\{\tilde{A}_c(\sigma)-\tilde{A}(\sigma)\}&=l'(\sum_{i=1}^n W_i^T\Sigma_i^* W_i+\Delta_\beta)^{-1}\Delta_{\beta}E(\tilde{\theta}-\theta)\\
&-d'\Omega^*U^T{\Sigma^*}^{-1}W(W^T{\Sigma^*}^{-1}W+\Delta_{\beta})^{-1}\Delta_{\beta}E(\tilde{\theta}-\theta).\\
&=-l'(\sum_{i=1}^n W_i^T\Sigma_i^* W_i+\Delta_\beta)^{-1}\Delta_{\beta}(\sum_{i=1}^n W_i^T\Sigma_i^* W_i+\Delta_\beta)^{-1}\Delta_\beta W_i\theta\\
&+d'\Omega^*U^T{\Sigma^*}^{-1}W(W^T{\Sigma^*}^{-1}W+\Delta_{\beta})^{-1}\Delta_{\beta}(\sum_{i=1}^n W_i^T\Sigma_i^* W_i+\Delta_\beta)^{-1}\Delta_\beta W_i\theta.
\end{align*}
It then can be checked that the order of the above bias is $O(n^{-2})$.

%\bigskip
\noindent {\bf Proof of Theorem \ref{thm3}:} To prove Theorem \ref{thm3}, we fist show (\ref{g1g2g3}).
Let $s'=m'\Omega^*U^T{\Sigma^*}^{-1}$. Then
\begin{align*}
\tilde{A}_c(\sigma)-A(\sigma)&=l'(W^T{\Sigma^*}^{-1}W+\pb)^{-1}W^TDY+m'\Omega^*U^T{\Sigma^*}^{-1}(Y-W\bb)\\
&+l'\wsw\pb\wsw W^T{\Sigma^*}^{-1}Y\\
&-s'W\wsw\pb\wsw W^T{\Sigma^*}^{-1}Y-l'b^{\beta}-m'b^{\nu}\\
&=s'(I-W\wsw W^T{\Sigma^*}^{-1})(U\bv+\varepsilon)\\
&+l'\wsw W^T{\Sigma^*}^{-1}(U\bv+\varepsilon)-m'b^{\nu}\\
&+s'W\wsw W^T{\Sigma^*}^{-1}(U\bv+\varepsilon).
\end{align*}
where ${\Sigma^*}=U_i(\Omega+P_{\nu_i})^{-1}U_i^{T}+R_i$ and $\Omega_i^*=(\Omega^{-1}+P_{\nu_i})^{-1}$.
%Since $\wsw(I+\pb\wsw)W^TDW=I-\wsw\pb\wsw\pb$ and $(I-W\wsw(I+\pb\wsw))W^TD)W=W\wsw\pb\wsw\pb$, we have
%\begin{align*}
%&m'V^{-1}U^TR^{-1}(I-W(W^TDW+\pb)^{-1}(I+\pb(W^TDW+\pb)^{-1})W^TD)W\bb\\
%&=m'V^{-1}U^TR^{-1}W(W^TDW+\pb)^{-1}\pb(W^TDW+\pb)^{-1})\pb\bb.
%\end{align*}
Let $m=(m_1',\cdots,m_n')'$. Then we have
$$
s'W=m'\Omega^{*}U^T{\Sigma^*}^{-1}W=(m_1'\Omega_1^{*}U_1^T{\Sigma^*}_1^{-1}W_1,\cdots,m_n'\Omega_n^{*}U_n^T{\Sigma^*}_n^{-1}W_n)^T.
$$
Thus if $m$ is sparse, then $s'W$ is also sparse. This implies that
$$s'W\wsw W^T{\Sigma^*}^{-1}=O(n^{-2}).$$
Therefore, we have
\begin{align*}
\tilde{t}_c(\sigma)-t(\sigma)&=l'(W^T{\Sigma^*}^{-1}W+\pb)^{-1}W^TDY+m'\Omega^*U^T{\Sigma^*}^{-1}(Y-W\bb)\\
&+l'\wsw\pb\wsw W^T{\Sigma^*}^{-1}Y\\
&-s'W\wsw\pb\wsw W^T{\Sigma^*}^{-1}Y-l'b^{\beta}-m'b^{\nu}\\
&=s'(I-W\wsw W^T{\Sigma^*}^{-1})(U\bv+\varepsilon)\\
&+l'\wsw W^T{\Sigma^*}^{-1}(U\bv+\varepsilon)-m'b^{\nu}+O_p(n^{-2})
\end{align*}
Using the matrix block inverse formula, we have
$$
{\Sigma^*}_i^{-1}={\Sigma}_i^{-1}+A_i
$$
where $A_i={\Sigma}_i^{-1} U_i \Omega (I+P_{\nu_i}\Omega-P_{\nu_i}\Omega U_i\Sigma^{-1} U_i \Omega P_{\nu_i})\Omega U_i {\Sigma}_i^{-1}$. Let $A=\mbox{diag}(A_1,\cdots,A_n)$. We then have
$$
\wsw=(W^T\Sigma^{-1}W)^{-1}-D
$$
where $D=(W^T\Sigma^{-1}W)^{-1}(I+B(W^T\Sigma^{-1}W)^{-1})^{-1} B (W^T\Sigma^{-1}W)^{-1}$ and $B=W^TAW+\pb$. It then follows that we have
\begin{align*}
\tilde{t}_c(\sigma)-t(\sigma)
&=s'(I-W((W^T\Sigma^{-1}W)^{-1}-D)W^T({\Sigma}^{-1}+A))(U\bv+\varepsilon)\\
&\quad+l'((W^T{\Sigma}^{-1}W)^{-1}-D) W^T({\Sigma}^{-1}+A)(U\bv+\varepsilon)-m'b^{\nu}+O_p(n^{-2})\\
&:=I_1+I_2+O_p(n^{-2})
\end{align*}
where
\begin{align*}
I_1&=s'(I-W(W^T\Sigma^{-1}W)^{-1}W^T{\Sigma}^{-1})(U\bv+\varepsilon)+l'(W^T{\Sigma}^{-1}W)^{-1} W^T{\Sigma}^{-1}(U\bv+\varepsilon)-m'b^{\nu},
\end{align*}
$I_2=(l'-s'W)\Delta_1(U\bv+\varepsilon)$
and $\Delta_1$ is defined before equation (\ref{g1g2g3}).  Using the above expression, we compute the second moment of $\tilde{t}_c(\sigma)-t(\sigma)$, which is equivalent to
\begin{align}
E\{(\tilde{A}_c(\sigma)-A(\sigma))^2\}&=(l-W^Ts)^T(W^T\Sigma^{-1}W)^{-1}(l-W^Ts)+m'(\Omega-\Omega U'\Sigma^{-1} U\Omega)d\nn\\
&+(l'-s'W)\Delta_1\Sigma\Delta_1^T(l'-s'W)^T\nn\\
&+2(l'-s'W)\Delta_1\Sigma(s'+(l'-s'W)(W^T\Sigma^{-1}W)^{-1}W^T\Sigma^{-1})^T\nn\\
&-2(l'-s'W)\Delta_1 U\Omega m'+O(n^{-2}).\label{result1}
\end{align}
This finishes the proof of (\ref{g1g2g3}).
%\bigskip

Let $\hat{\sigma}$ be the solution to (\ref{est:equa:variance}). Then it can be shown that
\begin{eqnarray}
MSE\{\hat{A}_c(\hat{\sigma})\}&=E\{(\hat{A}_c(\hat{\sigma})-A(\sigma))^2\}\nn\\
&=E\{(\hat{A}_c(\hat{\sigma})-\tilde{A}_c(\sigma))^2\}+E\{(\tilde{A}_c({\sigma})-A(\sigma))^2\}+O(n^{-2}).\label{mse:approxi}
\end{eqnarray}

Applying Taylor expansion on (\ref{est:equa:variance}), it can be shown that
\be
\label{sigma:expan}
\hat{\sigma}-\sigma=D^{-1}\tilde{e}+o_p(n^{-1/2})
\ee
where $D=\left(\frac{\partial \mathcal{Q}_{VC,k}(\sigma)}{\partial\sigma_l}\right)_{kl}$, $\tilde{e}=(e_1,e_2,\cdots,e_g)^T$ and $e_k=Y^TP\frac{\partial \Sigma^*}{\partial \sigma_k}PY-tr(P\Sigma P\frac{\partial \Sigma^*}{\partial \sigma_k})$.

Next, we want to show (\ref{mse:approxi}).  Note that
$$
\hat{A}_c(\hat\sigma)-A(\sigma)=\hat{A}_c(\hat\sigma)-\tilde{A}_c(\sigma)+\tilde{A}_c(\sigma)-A(\sigma)=\tilde{A}_c(\hat\sigma)-\tilde{A}_c(\sigma)+I_1+I_2.
$$
Also notice that $I_1=E\{\tilde{A}(\sigma)-A(\sigma)|Y\}$. It follows that
\begin{align*}
E\{(\tilde{A}_c(\hat\sigma)-A(\sigma))^2\}&=E\{(\hat{A}_c(\hat\sigma)-\tilde{A}_c(\sigma))^2\}+E\{(\tilde{A}_c(\sigma)-{A}(\sigma))^2\}\\
&\quad+2E\{(\hat{A}_c(\hat\sigma)-\tilde{A}_c(\sigma))I_1\}+2E\{(\hat{A}_c(\hat\sigma)-\tilde{A}_c(\sigma))I_2\}.
\end{align*}
Using the result in Kackar and Harville (1980), we have $E\{(\hat{A}_c(\hat\sigma)-\tilde{A}_c(\sigma))I_1\}=0$.  Moreover, by Cauchy-Swartcz inequality, we have
$$
E\{(\hat{A}_c(\hat\sigma)-\tilde{A}_c(\sigma))I_2\}\leq (E\{(\hat{A}_c(\hat\sigma)-\tilde{A}_c(\sigma))^2\})^{1/2}(E(I_2^2))^{1/2}.
$$
It can be checked that $E\{(\hat{A}_c(\hat\sigma)-\tilde{A}_c(\sigma))^2\}=O(n^{-1})$ and
\begin{align*}
E(I_2^2)&=(l-s'W)'(W^T\Sigma^{-1}W)^{-1}(I+\pb (W^T\Sigma^{-1}W)^{-1})^{-1}\\
&\quad +(W^T\Sigma^{-1}W)^{-1}(I+ (W^T\Sigma^{-1}W)^{-1}\pb)^{-1}(W^T\Sigma^{-1}W)^{-1}(l-W^T s)=O(n^{-3}).
\end{align*}
Hence, $E\{(\hat{A}_c(\hat\sigma)-\tilde{A}_c(\sigma))I_2\}=O(n^{-2})$. Therefore, we have
\begin{align}
E\{(\tilde{A}_c(\hat\sigma)-A(\sigma))^2\}&=E\{(\hat{A}_c(\hat\sigma)-\tilde{A}_c(\sigma))^2\}+E\{(\tilde{A}_c(\sigma)-{A}(\sigma))^2\}+O(n^{-2}). \label{result2}
\end{align}
This finishes the proof of (\ref{mse:approxi}).

Then
\bea
E\{(\hat{A}_c(\hat{\sigma})-\tilde{A}_c(\sigma))^2\}&=&2tr\{(BD^{-1}J\Sigma)^2\}+tr^2(\Sigma BD^{-1}J)+tr(D^{-1}B^T\Sigma BD^{-1}\Sigma_w)\nn\\
&&+4\sum_{j=1}^q\sum_{l=1}^q\lambda_j^T\Sigma(G_j\Sigma G_j+G_l\Sigma G_j)\Sigma\lambda_l+o(n^{-1})
\label{mse:2}
\eea
where $\Sigma_w=(2tr(G_i\Sigma G_j \Sigma))_{i,j}$.

%Lastly, we show (\ref{mse:2}).

To show (\ref{mse:2}), using the notation before Theorem 1, we have
$$
e_k=J_k^T(U\alpha+\epsilon)+(U\alpha+\epsilon)^TP \frac{\partial\Sigma^*}{\partial\sigma_k} P(U\alpha+\epsilon)-tr(P\Sigma P \frac{\partial\Sigma^*}{\partial\sigma_k})+o_p(n^{-1}).
$$
Let $
\bar{\mathbf{U}}_{\epsilon}=(\bar{U}_{\epsilon,1},\cdots,\bar{U}_{\epsilon,q})
$
where
$\bar{U}_{\epsilon,k}=(U\alpha+\epsilon)^TP \frac{\partial\Sigma^*}{\partial\sigma_k} P(U\alpha+\epsilon)-tr(P\Sigma P \frac{\partial\Sigma^*}{\partial\sigma_k})$
Then
\begin{align}
E\{(\hat{A}_c(\hat\sigma)-\tilde{A}_c(\sigma))^2\}:=V_1+V_2+V_3 \label{result3}
\end{align}
where $V_1=E\{[(U\alpha+\epsilon)^TBD^{-1}J(U\alpha+\epsilon)]^2\}$, $V_2=E\{[(U\alpha+\epsilon)^TBD^{-1}\bar{\mathbf{U}}_{\epsilon}]^2\}$ and $V_3=2E\{(U\alpha+\epsilon)^TBD^{-1}J(U\alpha+\epsilon)(U\alpha+\epsilon)^TBD^{-1}\bar{\mathbf{U}}_{\epsilon}\}$. Applying the standard result regarding the moments of quadratic forms of normally distributed random vectors, we have $V_3=0$ and
$$
V_1=2tr\{(BD^{-1}J\Sigma)^2\}+tr^2(\Sigma BD^{-1}J)
$$
and
$$
V_2=tr(D^{-1}B^T\Sigma BD^{-1}\Sigma_w)+4\sum_{j=1}^q\sum_{l=1}^q\lambda_j^T\Sigma(G_j\Sigma G_j+G_l\Sigma G_j)\Sigma\lambda_l.
$$
In summary of (\ref{result1}), (\ref{result2}) and (\ref{result3}), we conclude the result in Theorem 3. This finishes the proof of Theorem 3.

%\label{lastpage}

\end{document}